\documentclass{svproc}
\usepackage[utf8]{inputenc}

\usepackage{url}

\usepackage{natbib}
\bibpunct{(}{)}{;}{a}{}{,}%
\usepackage{graphicx}
\usepackage{amsmath,amssymb}
\usepackage{multirow}
\usepackage{longtable}

\usepackage[b5paper]{geometry}
\begin{document}
	\mainmatter              
	\title{Magneto-rotational evolution of neutron stars with hysteresis effect and fallback}
	\titlerunning{Magneto-rotational evolution of neutron stars}  
	%
	\author{Alena Khokhriakova\inst{1,2} \and Sergei Popov\inst{1,2}
		}
	\authorrunning{A. Khokhriakova and S. Popov} 
	%
	%
	\institute{Department of Physics, Lomonosov Moscow State University, \\Moscow, 119991 Russia,\\
		\and
		Sternberg Astronomical Institute, Lomonosov Moscow State University,\\ Moscow, 119234 Russia\\
			\email{alenahohryakova@yandex.ru}
}
	
	\maketitle

\begin{abstract}
	In recent years, accreting neutron stars (NSs) in X-ray binary systems in supernova remnants have been discovered. They are a puzzle for the standard magneto-rotational evolution of NSs, as their age ($\lesssim 10^5$ years) is much less than expected duration of the preceding Ejector and Propeller stages. 
	To explain such systems, we consider rotational evolution of NSs with fallback accretion and asymmetry in direct/backward transitions between Ejector and Propeller stages.
	It is shown that at certain values of the initial period and the magnetic field, a young neutron star may not enter the Ejector stage during its evolution.
\end{abstract}

\section{Evolution of neutron stars}

Recently, six accreting NSs in X-ray binary systems located in supernova remnants (SNRs) have been discovered, see \cite{2021arXiv210709325X}. They are specifically interesting objects  allowing to probe early evolution of accreting NSs, as in this cases the compact objects reach the stage of accretion very rapidly, in contradiction with standard assumptions (see e.g., the book \citealt{1992ans..book.....L}).
We discuss modifications in early evolution of NSs in close massive binaries.

\subsection{Hysteresis}

Transitions between evolutionary stages of  NS are determined by pressure balance at critical radii. However, this equality can be reached at different radii for direct and backwards cases.\footnote{We call ``direct'' transitions in the sequence Ejector~$\rightarrow$~Propeller~$\rightarrow$~Accretor.}
In particular, we are interested in asymmetry in Ejector-Propeller transition, first noticed by \cite{1970SvA....14..527S} and dubbed as hysteresis. 

In the phase of ejection the NS slows down. This results in decrease of the wind power. 
Direct transition Ejector~$\rightarrow$~Propeller happens when external pressure equalizes the relativistic wind pressure at gravitational capture radius ($R_\mathrm{G})$. 
Pressure of gravitationally captured cold matter grows towards the NS with distance as $r^{-5/2}$, i.e. faster than the wind pressure ($\sim r^{-2}$). So, particle generation and acceleration are stopped. 
 Transition in the opposite direction (Propeller$\rightarrow$Ejector) proceeds in a different way. When the propeller phase is on and accretion rate decreases, 
 the magnetospheric radius is growing gradually approaching the light cylinder,
where the pressure of external gravitationally captured matter is large. To overcome it, much larger wind power is needed than in the case of the equilibrium at $R_\mathrm{G}$. 

Here we apply this effect to the situation when a newborn NS experiences an intense fallback, and then while accretion rate ($\dot M$) is decreasing switches to Propeller, but can skip an Ejector phase due to hysteresis. Thus, time before wind accretion can be much shortened. 

\subsection{Fallback}

After formation of an NS, some amount of the progenitor star material expelled in the supernova explosion can fall back onto the compact object (\citealt{1989ApJ...346..847C}). 
We assume that in the case of fallback the NS is an Accretor when it starts to interact with the stellar wind.

If the star is isolated, fallback can last quite long. 
However, in a close binary system, expelled matter mostly leaves the Roche lobe of the future compact object.
Falling back it might feel joint gravity of both components. 
In addition, outside the Roche lobe the matter interacts with the stellar wind of the secondary component.
We assume that mostly matter which does not leave the Roche lobe of the compact object can fall back onto it. 
Thus, intensive fallback accretion cannot last long in binary systems, and so we can neglect this stage in our modeling, assuming as initial values of parameters those that they obtain after the fallback is over. 

After a short period of fallback accretion, the interaction of the NS and the companion star will be determined by the intensity of the stellar wind. 
Thus, the accretion rate would rapidly decrease.
Correspondingly, the size of the magnetosphere changes dramatically approaching the light cylinder. At this moment the  transition Propeller~$\rightarrow$~Ejector can happen. However, as it is not a direct transition, the hysteresis effect might be taken into account.
Such scenario was not considered in other works on this topic.

\section{Results}

We start our calculations when a brief period of fallback is over, and the external medium is determined by wind of the donor.
At this moment the NS has a period $p_0$, magnetic field $B_0$, and the accretion rate (from the wind) is $\dot{M}$.
If the NS does not become an Ejector at this moment, it will not enter this stage in the future (for non-decreasing rate of capture of external matter) and will start accreting quite fast if it is a Propeller.

\begin{figure}
	\includegraphics[width=\textwidth]{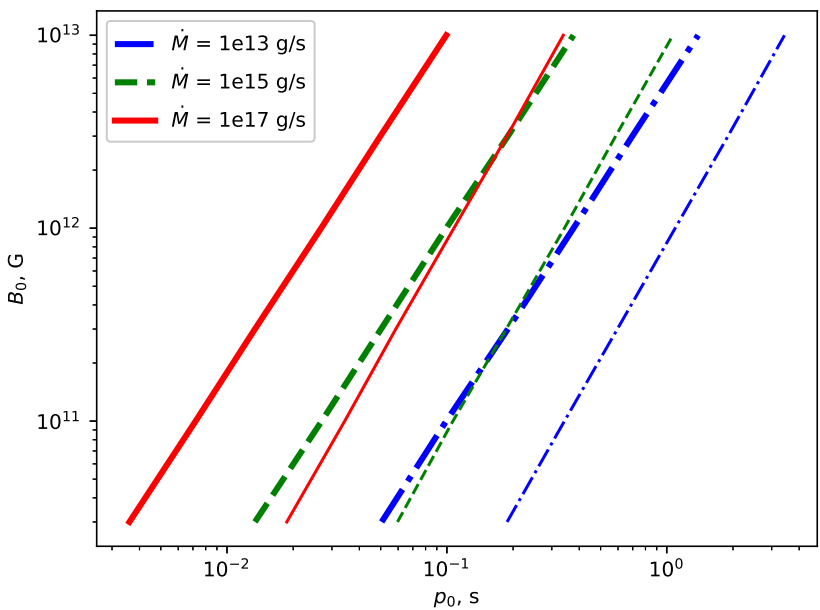}
    \caption{For three characteristic values of $\dot{M}$ we calculate critical values of $p_0$ and $B_0$. For a given $\dot{M}$, if the point is located to the left of the line, then a star is an Ejector after the intense fallback is over. If initial period is higher or magnetic field is lower than critical values, then an NS will not be in ejector phase.
Thick and thin lines represent the presence and absence of the hysteresis effect, respectively.}
    \label{fig:P0_B0}
\end{figure}

For three characteristic values of $\dot{M}$ we calculate critical values of $p_0$ and $B_0$ (Fig.~\ref{fig:P0_B0}). For a given $\dot{M}$, if the point is located to the left of the line, then a star is ``born'' as an Ejector (after short initial fallback accretion). If initial period is higher or magnetic field is lower than critical values, then an NS will not be in the ejector phase and will start to accrete quite fast. We compare our results with an approach without accounting for the hysteresis effect.
Bold and thin lines represent the presence and absence of the hysteresis effect, respectively.
The greater the accretion rate of the stellar wind, the easier it is to avoid the ejection phase.

Calculations for different sets of parameters ($p_0, B_0, \dot M$) allowed to reproduce situations when a NS reaches the stage of accretion within $10^5$~yrs, i.e. within the lifetime of a SNR.

\section{Conclusions}

 Presence of the fallback stage can change the evolution and observational properties of neutron stars in binary systems.
The hysteresis effect, proposed by Shvartsman, allows a young neutron star in a massive X-ray binaries to avoid the Ejector stage in the presence of fallback.
Such NSs can start accreting in a relatively short period, sometimes shorter than the lifetime of a SNR.  With  magnetic fields $\sim 10^{12}$~G and initial spin periods $\sim 0.1$~--~0.2~s NSs can avoid the Ejector stage if the accretion rate is $\gtrsim 10^{14}$~--~$10^{15}$~g~s$^{-1}$. 
Some of the massive binaries in SNRs can be examples of such systems.

\section*{Acknowledgements}
This work was supported by the Russian Science Foundation, grant 21-12-00141.  A.K. also acknowledges fellowship from ``Basis'' foundation, grant 20-2-1-77-1. 



\begin{thebibliography}{4}
\expandafter\ifx\csname natexlab\endcsname\relax\def\natexlab#1{#1}\fi

\bibitem[{{Chevalier}(1989)}]{1989ApJ...346..847C}
{Chevalier}, R.~A. 1989, \apj, 346, 847

\bibitem[{{Lipunov}(1992)}]{1992ans..book.....L}
{Lipunov}, V.~M. 1992, {Astrophysics of Neutron Stars}, {Springer-Verlag, Berlin Heidelberg}

\bibitem[{{Shvartsman}(1970)}]{1970SvA....14..527S}
{Shvartsman}, V.~F. 1970, \sovast, 14, 527

\bibitem[{{Xing} \& {Li}(2021)}]{2021arXiv210709325X}
{Xing}, Z.-p. \& {Li}, X.-d. 2021, arXiv e-prints, arXiv:2107.09325

\end{thebibliography}


\vskip 1cm

\begingroup              
\let\clearpage\relax

\endgroup

\end{document}